\documentclass[12pt,preprint]{aastex}
\usepackage{emulateapj5}

\newcommand\asca{{\it ASCA}}
\newcommand\sax{{\it BeppoSAX}}
\newcommand\chandra{{\it Chandra}}
\newcommand\rosat{{\it ROSAT}}

\newcommand\xmm{{\it XMM-Newton}}

\newcommand\kev{{\rm~keV}}
\newcommand\ev{{\rm~eV}}
\newcommand\kms{\ifmmode {\rm~km\ s}^{-1} \else ~km s$^{-1}$\fi}
\newcommand\Hunit{\ifmmode {\rm~km\ s}^{-1}\ {\rm Mpc}^{-1}
        \else ~km s$^{-1}$ Mpc$^{-1}$\fi}
\newcommand\ctssec{\ifmmode {\rm~count\ s}^{-1} \else ~count s$^{-1}$\fi}
\newcommand\ergsec{\ifmmode {\rm~erg\ s}^{-1} \else
        ~erg s$^{-1}$\fi}
\newcommand\funit{\ifmmode {\rm~erg\ s}^{-1}\;{\rm cm}^{-2} \else
        ~ergs s$^{-1}$ cm$^{-2}$\fi}
\newcommand\phflux{\ifmmode {\rm~photon\ s}^{-1}\;{\rm cm}^{-2}
        \else   ~photon s$^{-1}$ cm$^{-2}$\fi}
\newcommand\efluxA{\ifmmode {\rm~erg\ s}^{-1}\;{\rm cm}^{-2}\;{\rm
        \AA}^{-1} \else ~erg s$^{-1}$ cm$^{-2}$ \AA$^{-1}$\fi}
\newcommand\efluxHz{\ifmmode {\rm~erg\ s}^{-1}\;{\rm cm}^{-2}\;{\rm
        Hz}^{-1} \else ~erg s$^{-1}$ cm$^{-2}$ Hz$^{-1}$\fi}
\newcommand\cc{\ifmmode {\rm~cm}^{-3} \else cm$^{-3}$\fi}
\newcommand\FWHM{\ifmmode {\rm~FWHM} \else ${\rm~FWHM}$\fi}
\newcommand\Msun{\ifmmode M_{\odot} \else $M_{\odot}$\fi}
\newcommand\Lsun{\ifmmode L_{\odot} \else $L_{\odot}$\fi}

\newcommand\hbeta{\ifmmode {\rm H}\beta \else H$\beta$\fi}
\newcommand\Kalpha{\ifmmode {\rm K}\alpha \else K$\alpha$\fi}
\newcommand\NH{\ifmmode N_{\rm H} \else N$_{\rm H}$\fi}
\usepackage{graphicx}
\usepackage{here}

\begin{document}

\title{The Correlation between X-ray spectral slope and FeK$\alpha$ line energy in radio-quiet active galactic nuclei} 

\author{Gulab Chand Dewangan} 
\affil{Department of Astronomy \& Astrophysics, Tata Institute of Fundamental Research, Mumbai 400~005 India
 \\ 
{\tt gulab@tifr.res.in}}

\begin{abstract}
A significant correlation between FeK$\alpha$ line energy and X-ray spectral slope has been discovered among radio-quiet active galactic nuclei. The ionization stage of the bulk of the FeK$\alpha$ emitting material is not the same in all active galactic nuclei and 
is related to the shape of the X-ray continua.
Active galactic nuclei
with a steep X-ray spectrum tend to have a fluorescence FeK$\alpha$ line from 
highly ionized material. In the narrow-line Seyfert~1 galaxies with 
steeper X-ray spectrum ($\Gamma_X \ga 2.1$), the FeK$\alpha$ line originates from highly ionized 
material. In the Seyfert~1 galaxies and quasars with flatter X-ray spectrum 
($\Gamma_X \la 2.1$), bulk of
the FeK$\alpha$ emission arises from near neutral or weakly ionized material.
The correlation is an important observational characteristic related to the accretion process in radio quiet active galactic nuclei and is driven by a fundamental physical parameter which is likely to be the accretion rate relative to the Eddington rate.
\end{abstract}


\section{Introduction}
X-ray emission of active galactic nuclei (AGNs) consists of a power-law with a photon index of $\sim 1.9$ \citep{Mushotzky97}, a soft X-ray excess component at lowest X-ray energies \citep{SGN85,Arnaudetal85}, a strong emission line from the K-shell of iron at $\sim 6.4\kev$ \citep[see e.g.,][]{Tanakaetal95}, and the Compton-reflection hump in the energy range of $\sim 20 - 100\kev$ 
and peaking at $\sim 30\kev$ \citep{GF91,Reynolds99}. This basic form of the X-ray spectrum is easily explained in the framework of disk-corona models \citep[e.g.,][]{HM93}, where an optically-thin corona irradiates a dense and thin accretion disk surrounding a super-massive black hole (SMBH). The optically thick accretion disk is often assumed to be weakly ionized and radiatively efficient therefore cold. 
The broad FeK$\alpha$ line at $\sim6.4\kev$ 
arises due to fluorescence of neutral iron in the inner regions of the accretion disk \citep{RF97}. 
%
The FeK$\alpha$ line and the Compton reflection both arise due to irradiation of coronal hard X-ray emission onto the disk suggesting that any change in the FeK$\alpha$ line emission, due to some change in the accretion disk or corona, must be accompanied by the corresponding change in the reflection hump and therefore the observed shape of the X-ray continua. The energy of the FeK$\alpha$ emission depends on the ionization stage of iron and ranges from $6.4\kev$ for Fe~I to $6.9\kev$ for Fe~XXVI. Indeed the rest-frame energy of the K$\alpha$ emission from different AGNs has been observed to cover a range of $\sim 6.4-6.9\kev$ (see Table~\ref{tab1}). Also the photon index of the X-ray spectra of AGNs ranges from $\sim 1.7$ to $\sim2.5$. Hence it is of great importance to investigate whether the above two parameters are related. 

\section{The data}
The $2-10\kev$ photon indices, the rest-frame energies of FeK$\alpha$ 
emission, and the width of the $\hbeta$ line of 32 AGNs were obtained from the published literature and are listed in Table~\ref{tab1}.  
For most of the 
AGNs which were observed with \asca, the peak energy of the FeK$\alpha$ was 
derived by using a single Gaussian.  Some AGNs, observed with \xmm, showed 
clear evidence for the presence of both the narrow and broad components 
of the FeK$\alpha$ emission. For these AGNs, line energies of narrow and 
broad components are listed in Table~\ref{tab1}. All the AGNs listed in 
Table~\ref{tab1} show strong FeK$\alpha$ emission. These AGNs were 
selected based on the following criteria: ($i$) classified as a radio-quiet and a type~1 AGN, ($ii$)  the equivalent width of the FeK$\alpha$ line (EW$_{\rm FeK\alpha}$) is 
$\ga100\ev$ or the line has been resolved to be broad (FWHM $>10000\kms$) or the line is rapidly variable, and ($iii$) the rest-frame energy of the FeK$\alpha$ line has been determined with an accuracy better than $5\%$. The second criterion is important to 
exclude AGNs which show only the narrow unresolved FeK$\alpha$ line arising outside of the accretion disk. It is thought that the 
unresolved narrow component of the FeK$\alpha$ line with equivalent widths 
$< 100\ev$ is unlikely to arise from the inner regions of the accretion disk \citep{Reynolds01}. 
Such narrow unresolved components have been observed with \chandra and \xmm 
from a number of AGNs \citep[e.g.,][]{Kaspietal01,Poundsetal01,Turneretal02}, 
and are found to have their peak energy at $6.4\kev$ with equivalent width of 
$\sim 50-100\ev$. These AGNs are not part of the sample listed in Table~\ref{tab1}. 
Here the aim is to study the 
FeK$\alpha$ emission originating from the accretion disk.
\begin{table*}[t]
\begin{center}
\caption{X-ray spectral slope, rest-frame energy of FeK$\alpha$ emission, and width of $\hbeta$ line for 32 radio-quiet AGNs \label{tab1}}
{\scriptsize
\begin{tabular}{lcccccccc}
\tableline\tableline
Object & $\Gamma_{\rm X}$ & \multicolumn{3}{c}{$E_{\rm FeK\alpha}$\tablenotemark{a}} & Model\tablenotemark{b} & Source of  & FWHM$_{{\rm H}\beta}$  &   Ref.
\\
       & ($2-10\kev$) & single & narrow & broad &  & data &  ($\kms$  & \\
              &              & ($\kev$) & ($\kev$) & ($\kev$) & &  &  &  \\
\tableline
I~Zw~1   &  $2.37_{-0.05}^{+0.05}$ & $6.77_{-0.17}^{+0.11}$ & -- & -- &PL+GA&\asca  &  1250 & 1,24 \\
PHL~909 &  $1.11_{-0.11}^{+0.11}$ & $6.40_{-0.22}^{+0.18}$ & -- & -- &PL+GA & \asca  & 11000 & 1,25 \\
HE~1029$-$1401 & $1.83_{-0.05}^{+0.05}$ & $6.62_{-0.19}^{+0.16}$ & -- & -- &PL+GA & \asca  & 7500 &  1,26 \\
PG~1114$+$445  & $1.71_{-0.06}^{+0.06}$ & $6.43_{-0.06}^{+0.06}$ & -- & -- &PL+GA & \asca & 4570  & 1,24 \\
PG~1116$+$215  & $2.09_{-0.05}^{+0.05}$ & $6.76_{-0.08}^{+0.08}$ & -- & -- &PL+GA & \asca  &  2920 & 1,24 \\
PG~1211$+$143    & $2.06_{-0.05}^{+0.05}$ & $6.41_{-0.16}^{+0.09}$ & -- & -- &PL+GA &\asca  & 1860 & 1,24 \\
Mrk~205        & $1.80_{-0.04}^{+0.04}$ & -- & $6.39_{-0.04}^{+0.04}$ & $6.67_{-0.10}^{+0.10}$ &PL+GA+DL & \xmm  & 3150 &  2,25 \\
RBS~1259  & $2.36_{-0.05}^{+0.05}$ & $6.76_{-0.21}^{+0.29}$ & -- & -- &PL+GA & \asca  & 2200 &  1,27 \\
PG~1416$-$129    & $1.78_{-0.02}^{+0.02}$ & $6.54_{-0.18}^{+0.16}$ & -- & -- &PL+GA &\asca  & 6110 & 1,24 \\
NGC~3516       & $1.56_{-0.04}^{+0.04}$ & $6.40_{-0.07}^{+0.00}$ & -- & -- &PL+DL &\asca  & 4760 & 3,28 \\
NGC~4051       & $1.84_{-0.03}^{+0.05}$ & $6.30_{-0.24}^{+0.24}$ & -- & -- &PL+GA & \asca  & 1120 & 5,29 \\
PG~1501$+$106      & $1.60_{-0.04}^{+0.03}$ & $6.38_{-0.23}^{+0.17}$ & -- & -- &PL+GA & \asca & 5470 & 6,24 \\
PG~1534$+$58       & $1.68_{-0.03}^{+0.02}$ & $6.34_{-0.19}^{+0.16}$ & -- & -- &PL+GA & \asca  & 5340 & 6,24 \\
Mrk~359        & $1.85_{-0.04}^{+0.04}$ & $6.43_{-0.03}^{+0.03}$ & -- & -- &PL+GA & \xmm  & 480 & 7,30 \\
Mrk~335\tablenotemark{c}   & $2.29_{-0.02}^{+0.02}$ & $6.97$ & -- & -- &PL+DL & \xmm  & 1720  & 8,28 \\
E~1821$+$643     & $1.76_{-0.05}^{+0.05}$ & $6.49_{-0.06}^{+0.06}$ & -- & -- &PL+DL & \chandra & 5900   & 12,31 \\
MCG-6-30-15 & $1.94_{-0.02}^{+0.02}$ & $6.95_{-0.15}^{+0.00}$ & -- & -- &IR+DL & \xmm  & 2400  & 13,32 \\
Mrk~279        & $2.03_{-0.02}^{+0.02}$ & $6.50_{-0.08}^{+0.08}$ & -- & -- &PL+DL & \asca  & 6200 & 14.33 \\
Mrk~841 & $2.02_{-0.04}^{+0.06}$ & $6.41_{-0.06}^{+0.05}$ & -- & -- & PL+GA & \xmm & 5470 & 15,24 \\
Akn~564        & $2.538_{-0.005}^{+0.005}$ & $6.99_{-0.13}^{+0.01}$ & -- & -- &PL+DL &\asca & 865 & 16,29 \\
Ton~S180       & $2.44_{-0.02}^{+0.02}$ & $6.71_{-0.14}^{+0.12}$ & -- & -- &PL+GA &\asca  & 1085 & 17,29 \\
Fairall~9      & $1.73_{-0.07}^{+0.07}$ & $6.38_{-0.03}^{+0.03}$ & -- & --& PL+GA & \xmm & 6500 & 18,34 \\
PG~0947+396    & $1.95_{-0.10}^{+0.10}$ & $6.35_{-0.13}^{+0.13}$ & -- & -- & PL+GA & \sax & 4830 & 19,24 \\
PG~1115+407    & $2.40_{-0.13}^{+0.13}$ & $6.69_{-0.11}^{+0.11}$ & -- & -- & PL+GA & \sax & 1720 & 19,24 \\
PG~1352+183    & $2.30_{-0.16}^{+0.16}$ & $6.43_{-0.16}^{+0.16}$ & -- & -- & PL+GA & \sax & 3600 & 19,24 \\
PG~1244+026    &$2.35_{-0.10}^{+0.10}$  & $7.0_{-0.10}^{+0.10}$  & -- & -- & PL+GA & \asca & 830 & 20.24 \\
Mrk~766        & $2.09_{-0.02}^{+0.02}$ & $6.6_{-0.1}^{+0.1}$ & -- & -- & PL+DL & \xmm & 850 & 21,27   \\
Mrk~478     & $1.98_{-0.03}^{+0.03}$ & $6.37_{0.07}^{+0.07}$  & -- & --  & PL+GA& \asca & 1450  & 1,24 \\
RX~J0148-27 & $1.99_{-0.17}^{+0.17}$ & $6.5_{-0.2}^{+0.2}$ & -- & -- & PL+GA & \asca & 1050  & 20 \\
NGC~3783 & $1.60_{-0.02}^{+0.02}$ & -- & $6.40_{-0.01}^{+0.01}$ & $6.29_{-0.03}^{+0.03}$ & PL+2GA & \xmm & 5672 & 22,35 \\
MCG-2-58-22 & $1.58_{-0.08}^{+0.08}$ & $6.33_{-0.07}^{0.18}$ & -- & -- & PL+GA & \asca & 6360 & 3,30 \\
NGC~7469 & $1.78_{-0.07}^{+0.07}$ & $6.39_{-0.05}^{+0.06}$ & -- & -- & PL+GA & \asca & 3460  & 3,28  \\
\tableline
\end{tabular}}
\tablenotetext{a}{Rest-frame central energy of the FeK$\alpha$ line}
\tablenotetext{b}{Spectral model used to derive the photon index and iron line energy. PL = power-law continuum, GA = Gaussian line profile, DL = disk-line model of either \citet{Fabianetal89} or \citet{Laor91}}
\tablenotetext{c}{The central energy of FeK$\alpha$ line was fixed in the disk-line model. The Laor model gave a best-fit energy of $7.2\pm0.2\kev$ \citep{Gondoinetal02}}
\tablerefs{
(1) \citet{RT00}; (2) \citet{Reevesetal01}; (3) \citet{Nandraetal99};
(4) \citet{DMZ00}; (5) \citet{Guainazzietal96}; (6) \citet{Georgeetal00}; (7) \citet{OBrienetal01};
(8) \citet{Gondoinetal02}; (9) \citet{Wangetal99}; (10) \citet{Mattetal01}; 
(11) \citet{Yaqoobetal96}; (12) \citet{Fangetal02}; 
(13) \citet{Wilmsetal01}; (14) \citet{WGY01}; (15) \citet{Petruccietal02}; 
(16) \citet{Turneretal01}; (17) \citet{Romanoetal02}; (18) \citet{Gondoinetal01}; 
(19) \citet{Mineoetal00}; (20) \citet{Vaughanetal99}; (21) \citet{Pageetal01}; (22) \citet{Blustinetal02}; (23) \citet{Salvietal02}; (24) \citet{BG92}; (25) \citet{ZO90}; (26) \citet{WWR91}; (27) \citet{Grupeetal98}; (28) \citet{Crenshaw86}; (29) \citet{VVG01}; 
(30) \citet{OS82}; (31) \citet{Corbin91}; (32) \citet{Winkler92}; (33) \citet{Osterbrock77};
 (34) \citet{HP78}; (35) \citet{Evans88} }
\end{center}
\end{table*}

\section{The Correlation} 
\begin{figure*}
\centering
\includegraphics[width=4.6cm]{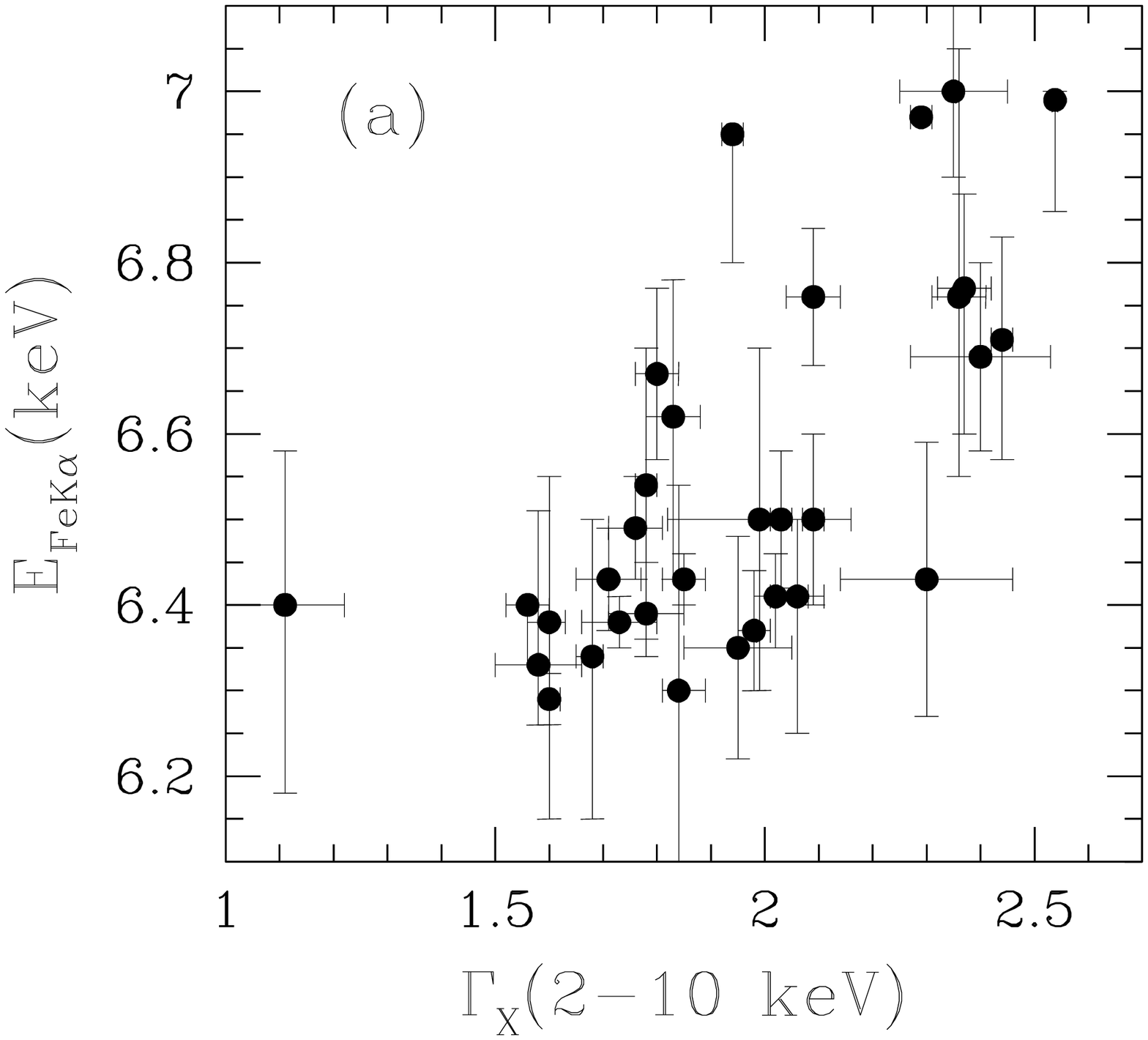}
\includegraphics[width=4.6cm]{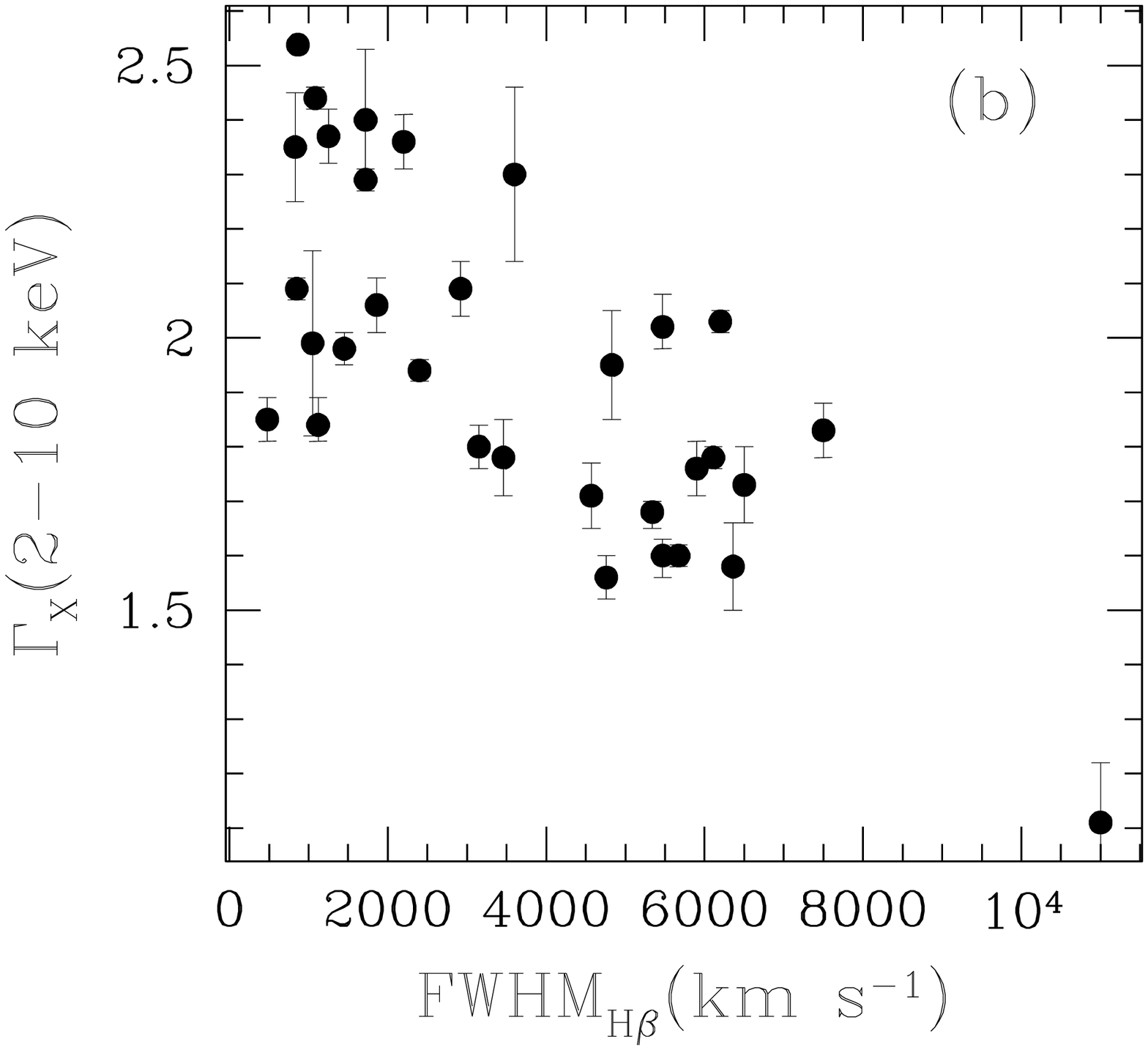}
\includegraphics[width=4.6cm]{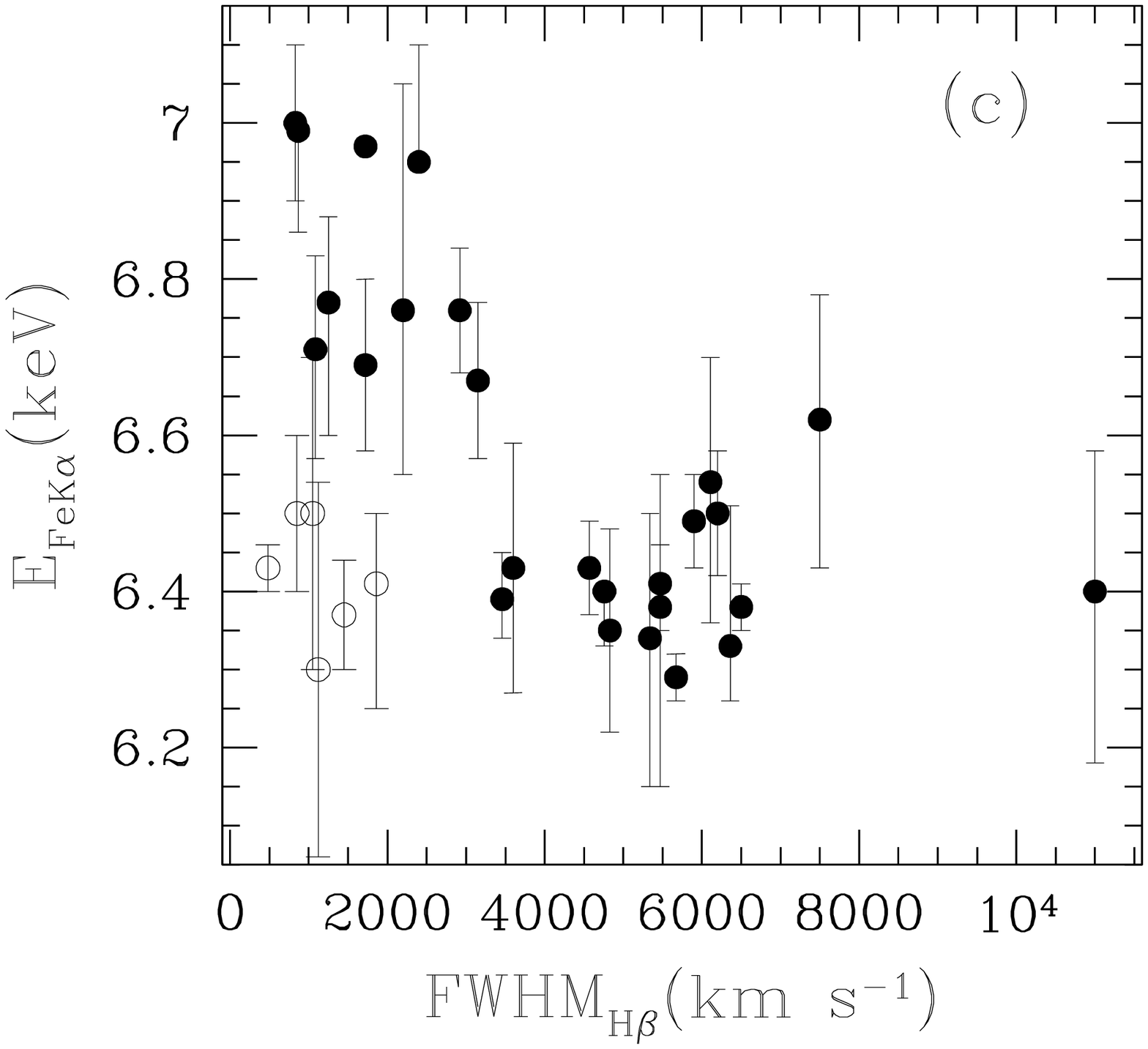}
\caption{(a) The $2-10\kev$ power-law photon index ($\Gamma_X$) plotted against the rest-frame line energy of the FeK$\alpha$ emission ($E_{FeK\alpha}$) for the 32 AGNs listed in Table~\ref{tab1}. A strong correlation between $\Gamma_X$ and $E_{FeK\alpha}$ is evident.
(b) The $2-10\kev$ power-law photon index ($\Gamma_X$) plotted against FWHM of the $\hbeta$ line. NLS1 galaxies occupy the region with FWHM$_{{\rm H}\beta}\la2000\kms$.
(c) The rest-frame energy of the FeK$\alpha$ line plotted against the FWHM of the $\hbeta$ line. NLS1 galaxies with $\Gamma_X < 2.1$ are shown as open circles.}
\label{f1}
\end{figure*}

Figure~\ref{f1}(a) shows the plot of the power-law photon index against 
the rest-frame line energy of the FeK$\alpha$ emission for the AGNs listed 
in Table~\ref{tab1}. For AGNs which show narrow as well as broad components of FeK$\alpha$ line, the line energies plotted are the rest-frame energy of the broad components which were derived by using a Gaussian or disk-line model of \citet{Fabianetal89} or \citet{Laor91}. Some AGNs show strong red wings in their FeK$\alpha$ line profile due to relativistic effects. Such redshifted lines are not fitted by a Gaussian but are well described by disk-line models of \citet{Fabianetal89} or \citet{Laor91}. In such cases, the appropriate rest-frame energy of the FeK$\alpha$ line is that derived from the disk-line models.

As can be seen in Fig.~\ref{f1}(a), the $2-10\kev$ X-ray 
photon index appears to be correlated with the rest-frame energy of 
the FeK$\alpha$ line. In order to quantify this correlation, the 
Spearman's rank-correlation coefficient ($\rho$) was calculated.
For the sample of 32 AGNs, $\rho$ is found to be $0.69$ for $30$ degrees of 
freedom. This correlation is statistically significant at a level of $>99.9\%$ 
as inferred from the Student's t-test.

The $2-10\kev$ photon index is plotted against the FWHM of the $\hbeta$ line in Figure~\ref{f1}(b) for the sample listed in Table~\ref{tab1}. 
The two parameters are strongly anti-correlated ($\rho=-0.67$) with a significance 
level $>99.9\%$. This correlation is the same as that reported by \citet{BME97} and verifies their result. In Fig.~\ref{f1}(b), AGNs with FWHM$_{\hbeta} \la 2000\kms$ are the narrow-line Seyfert~1 galaxies(NLS1; \citealt{OP85}). Some NLS1s e.g., NGC~4051, Mrk~359, Mrk~478, have photon indices similar to that of broad-line Seyfert~1 galaxies (BLS1s; see \citealt{Dewanganetal02} for the definition).

In Figure~\ref{f1}(c), the rest-frame energy of the FeK$\alpha$ emission is plotted 
against the width of the $\hbeta$ line. NLS1 galaxies show a large diversity in their FeK$\alpha$ line energies. However, the FeK$\alpha$ emission arises from near neutral material only from those NLS1 galaxies with flatter X-ray spectrum ($\Gamma_X < 2.1$) similar to that of BLS1 galaxies. The energy of the FeK$\alpha$ emission appears to be anti-correlated with the width of $\hbeta$ line ($\rho=-0.44$) at a significance level of $98.6\%$. Excluding the 6 NLS1 galaxies with $\Gamma_X < 2.1$, the correlation improves to a significance level of $99\%$ ($\rho = -0.71$).
\section{Discussion}
A significant correlation between the $2-10\kev$ photon index and the rest-frame energy
of the FeK$\alpha$ line has been discovered in radio-quiet AGNs. This correlation adds
to the many correlations found earlier in AGNs. \citet{BG92} found numerous correlations
between optical line widths and strengths. Most notably was the correlation between the width of the $\hbeta$ line and strength of Fe~II emission. They identified a driving parameter ``eigen vector 1'' through principal component analysis. \citet{BBF96} discovered  the correlation between the $0.1-2.4\kev$ photon index and the width of the $\hbeta$ line. 
\citet{BME97} reported that the \asca $2-10\kev$ photon index and the width of the $\hbeta$
line are also correlated. Further, \rosat and \asca studies have confirmed that Seyfert~1s with steeper power-law index appear to be more variable on shorter time scales \citep{KSW97,Fioreetal98,Turneretal99}. These correlations involving X-ray and optical observations  suggest that the same eigen vector 1 is responsible for the width of the $\hbeta$ line, strength of the FeII emission, shape of the X-ray continuum, X-ray variability, and the energy of the FeK$\alpha$ line. 
The eigen vector 1 is identified as the accretion rate relative to the Eddington rate ($\dot{m}=\frac{\dot{M}}{\dot{M}_{Edd}}$) \citep[see e.g.,][]{PDO95,Brandt00}. This implies that the extreme line energy of the FeK$\alpha$ emission of AGNs is also due to a higher fractional accretion rate $\dot{m}$.

NLS1 galaxies show a large diversity in their X-ray properties (Fig.~\ref{f1}; \citealt{BBF96}; \citealt{BME97}) and can be divided into two groups --  ($i$) NLS1s with X-ray properties similar to that of BLS1s e.g., Mrk~359, and ($ii$) NLS1s with extreme X-ray properties -- steeper spectrum, FeK$\alpha$ emission from highly ionized material e.g., Akn~564. NLS1s in the former group are likely to have smaller black hole masses but similar accretion rates as that of BLS1s. The exceptionally narrow (FWHM$\sim100\kms$) forbidden lines, FWHM$_{\hbeta}\sim500\kms$, $\Gamma_X\sim1.85$, and $E_{FeK\alpha}\sim6.4$ of Mrk~359 favor such a scenario. NLS1s with extreme X-ray properties are probably the AGNs with Eddington or super-Eddington accretion rates and/or smaller black hole masses.

It is unlikely that the bulk of the FeK$\alpha$ emitting material is moving in  a preferred direction suggesting that Doppler effect is not responsible for the observed range of $\sim6.4-7\kev$ of E$_{FeK\alpha}$. For a given emissivity law and location of the FeK$\alpha$ emitting material, the diversity in the line profile, due to gravitational effects, is limited by the range of inclination angles. For type~1 AGNs, gravitational effects may result in the observed range of line energies if the FeK$\alpha$ emitting material is highly ionized but its location with respect to the SMBH varies substantially among AGNs. In this case, the disk-line fits to the FeK$\alpha$ line should always result in the rest-frame line energy of $\sim6.9\kev$ which is not the case (see Table~\ref{tab1}). The energy of the FeK$\alpha$ line depends on the ionization stage of iron and the observed range could easily be produced by neutral to H-like iron. 
The straight forward interpretation of the correlation between $\Gamma_X$ and $E_{FeK\alpha}$ is that the accretion disks of AGNs with steeper X-ray continua are more ionized. Observations of H-like oxygen, nitrogen, carbon emission lines from MCG-6-30-15 and Mrk~766 \citep{BRetal01} supports the above picture. NLS1s with steep X-ray spectrum seem to posses highly ionized accretion disks. In the disk-corona models, the $2-10\kev$ continuum is thought to arise from the corona, while the FeK$\alpha$ line arises due to fluorescence in the accretion disk. The question then is how the ionization state of the accretion disk affects the slope of the X-ray continuum. 

A steeper primary X-ray continuum can be produced by comptonization of soft photons in a cooler corona \citep{PDO95}. If the observed $2-10\kev$ continuum of AGNs is dominated by the primary emission, then the observed correlation between $\Gamma_X$ and $E_{FeK\alpha}$ imply that cooler coronae are associated with ionized accretion disks. It is known that BLS1s and NLS1s with steep X-ray spectrum have comparable X-ray luminosities. If the geometry of the disk-corona systems in the two types of Seyfert~1s is similar, it is unlikely that photo-ionization of the disk material will result in different stages of ionizations. There can be other processes e.g., thermal ionization in the inner regions of a disk with high accretion rates. Such disks emit soft X-ray excess emission which can maintain the corona at a lower temperature \citep{PDO95}.

Under certain conditions, observed $2-10\kev$ continuum can be dominated by the Compton reflection component.
The strength and the shape of the reflection component depends on the solid angle
subtended by the accretion disk onto the corona and ionization state of the reflector.  The spectral shape of the reflection component depends on the competition between the two physical processes -- ($i$) photo-electric absorption of the coronal photons by the atoms in the accretion disk, and ($ii$) Compton scattering of coronal photons by the electrons  in the disk. If the disk is fully ionized then the Compton scattering dominates over the photo-electric absorption. 
Under these conditions, hard as well as soft X-ray photons are  scattered and the reflection hump is expected to extend to the soft X-ray regime. 
If the accretion rate is high, it is likely that the disk is radiation pressure dominated \citep{Fabianetal02}. Under these conditions, the disk can be clumpy, irregular or ribbed \citep[see][]{LE74,GR88}. If the corona lies close to the disk, the solid angle subtended by the disk onto the corona can be much larger than $2\pi$ and coronal photons can be reflected many times before they escape. The net effect is that the resulting spectrum is dominated by reflection and considerably steep \citep{RFB02}. 
Thus a highly ionized irregular disk could give rise to both the steep X-ray spectrum and the K$\alpha$ emission from highly ionized iron. Both of these conditions are most likely to occur when the accretion rate is high and close to the Eddington rate.

\section{Conclusions}
A correlation between the photon index of the $2-10\kev$ X-ray continua and the central energy of the FeK$\alpha$ emission has been discovered for the radio quiet AGNs. This correlation suggests that AGNs with steep soft X-ray continua have highly ionized accretion disks. 
\acknowledgements
Prof. A. R. Rao is gratefully acknowledged for the discussions and his suggestions on this paper. The author is thankful to the referee of this paper, Dr. Ehud Behar, for important suggestions and comments that improved the paper. The author acknowledges the partial support from the Kanwal Rekhi scholarship program of the TIFR Endowment Fund.


\begin{thebibliography}{}

\bibitem[Arnaud et al.(1985)]{Arnaudetal85} Arnaud, K.~A.~et al.\ 
1985, \mnras, 217, 105 

\bibitem[{{Boller} {et~al.}(1996){Boller}, {Brandt}, \& {Fink}}]{BBF96}
{Boller}, T., {Brandt}, W.~N., \& {Fink}, H. 1996, \aap, 305, 53+

\bibitem[Blustin et al.(2002)]{Blustinetal02} Blustin, A.~J., 
Branduardi-Raymont, G., Behar, E., Kaastra, J.~S., Kahn, S.~M., Page, 
M.~J., Sako, M., \& Steenbrugge, K.~C.\ 2002, \aap, 392, 453 

\bibitem[{{Boroson} \& {Green}(1992)}]{BG92}
{Boroson}, T.~A. \& {Green}, R.~F. 1992, \apjs, 80, 109

\bibitem[{{Brandt} {et~al.}(1997){Brandt}, {Mathur}, \& {Elvis}}]{BME97}
{Brandt}, W.~N., {Mathur}, S., \& {Elvis}, M. 1997, \mnras, 285, L25

\bibitem[Brandt(2000)]{Brandt00} Brandt, W.~N.\ 2000, IAU 
Symp.~195: Highly Energetic Physical Processes and Mechanisms for Emission 
from Astrophysical Plasmas, 195, 207 


\bibitem[Branduardi-Raymont et al.(2001)]{BRetal01} 
Branduardi-Raymont, G., Sako, M., Kahn, S.~M., Brinkman, A.~C., Kaastra, 
J.~S., \& Page, M.~J.\ 2001, \aap, 365, L140 

\bibitem[Crenshaw(1986)]{Crenshaw86} Crenshaw, D.~M.\ 1986, \apjs, 
62, 821 


\bibitem[{{Dewangan} {et~al.}(2002){Dewangan}, {Boller}, {Singh}, \&
  {Leighly}}]{Dewanganetal02}
{Dewangan}, G.~C., {Boller}, T., {Singh}, K.~P., \& {Leighly}, K.~M. 2002,
  \aap, 390, 65

\bibitem[Done, Madejski, {\. Z}ycki(2000)]{DMZ00} Done, 
C., Madejski, G.~M., {\. Z}ycki, P.~T.\ 2000, \apj, 536, 213 

\bibitem[Evans(1988)]{Evans88} Evans, I.~N.\ 1988, \apjs, 67, 
373 


\bibitem[Fabian et al.(1989)]{Fabianetal89} Fabian, 
A.~C., Rees, M.~J., Stella, L., \& White, N.~E.\ 1989, \mnras, 238, 729 



\bibitem[{{Fabian} {et~al.}(2002){Fabian}, {Ballantyne}, {Merloni}, {Vaughan},
  {Iwasawa}, \& {Boller}}]{Fabianetal02}
{Fabian}, A.~C., {Ballantyne}, D.~R., {Merloni}, A., {Vaughan}, S., {Iwasawa},
  K., \& {Boller}, T. 2002, \mnras, 331, L35

\bibitem[Fang et al.(2002)]{Fangetal02} Fang, T., Davis, D.~S., 
Lee, J.~C., Marshall, H.~L., Bryan, G.~L., \& Canizares, C.~R.\ 2002, \apj, 
565, 86 


\bibitem[{{Fiore} {et~al.}(1998){Fiore}, {Laor}, {Elvis}, {Nicastro}, \&
  {Giallongo}}]{Fioreetal98}
{Fiore}, F., {Laor}, A., {Elvis}, M., {Nicastro}, F., \& {Giallongo}, E. 1998,
  \apj, 503, 607

\bibitem[{{George} \& {Fabian}(1991)}]{GF91}
{George}, I.~M. \& {Fabian}, A.~C. 1991, \mnras, 249, 352

\bibitem[George et al.(2000)]{Georgeetal00} George, I.~M., Turner, 
T.~J., Yaqoob, T., Netzer, H., Laor, A., Mushotzky, R.~F., Nandra, K., \& 
Takahashi, T.\ 2000, \apj, 531, 52 

\bibitem[Gondoin et al.(2001)]{Gondoinetal01} Gondoin, P., Lumb, D., 
Siddiqui, H., Guainazzi, M., \& Schartel, N.\ 2001, \aap, 373, 805 



\bibitem[Gondoin et al.(2002)]{Gondoinetal02} 
Gondoin, P., Orr, A., Lumb, D., \& Santos-Lleo, M.\ 2002, \aap, 388, 74 

\bibitem[Grupe et al.(1998)]{Grupeetal98} 
Grupe, D., Wills, B.~J., Wills, D., \& Beuermann, K.\ 1998, \aap, 333, 827 


\bibitem[Guainazzi et al.(1996)]{Guainazzietal96} 
Guainazzi, M., Mihara, T., Otani, C., \& Matsuoka, M.\ 1996, \pasj, 48, 781 


\bibitem[{{Guilbert} \& {Rees}(1988)}]{GR88}
{Guilbert}, P.~W. \& {Rees}, M.~J. 1988, \mnras, 233, 475

\bibitem[{{Haardt} \& {Maraschi}(1993)}]{HM93}
{Haardt}, F. \& {Maraschi}, L. 1993, \apj, 413, 507

\bibitem[Hawley \& Phillips(1978)]{HP78} Hawley, S.~A.~\& Phillips, M.~M.\ 1978, \apj, 225, 780 


\bibitem[Kaspi et al.(2001)]{Kaspietal01} Kaspi, S.~et al.\ 2001,
\apj, 554, 216

\bibitem[{{Koenig} {et~al.}(1997){Koenig}, {Staubert}, \& {Wilms}}]{KSW97}
{Koenig}, M., {Staubert}, R., \& {Wilms}, J. 1997, \aap, 326, L25





\bibitem[Laor(1991)]{Laor91} Laor, A.\ 1991, \apj, 376, 90 


\bibitem[{{Lightman} \& {Eardley}(1974)}]{LE74}
{Lightman}, A.~P. \& {Eardley}, D.~M. 1974, \apjl, 187, L1

\bibitem[Matt et al.(2001)]{Mattetal01} Matt, G., Guainazzi, M., 
Perola, G.~C., Fiore, F., Nicastro, F., Cappi, M., \& Piro, L.\ 2001, \aap, 
377, L31 

\bibitem[Mineo et al.(2000)]{Mineoetal00} Mineo, T.~et al.\ 2000, 
\aap, 359, 471 


Tanaka, Y.\ 1995, \mnras, 272, L9 

\bibitem[Mushotzky(1997)]{Mushotzky97} Mushotzky, R.~F.\ 1997, ASP 
Conf.~Ser.~128: Mass Ejection from Active Galactic Nuclei, 141 




\bibitem[Nandra et al.(1999)]{Nandraetal99} Nandra, K., George, 
I.~M., Mushotzky, R.~F., Turner, T.~J., \& Yaqoob, T.\ 1999, \apjl, 523, L17 

\bibitem[O'Brien et al.(2001)]{OBrienetal01} O'Brien, P.~T., Page, 
K., Reeves, J.~N., Pounds, K., Turner, M.~J.~L., \& Puchnarewicz, E.~M.\ 
2001, \mnras, 327, L37 

\bibitem[Osterbrock(1977)]{Osterbrock77} Osterbrock, D.~E.\ 1977, 
\apj, 215, 733 

\bibitem[{{Osterbrock} \& {Pogge}(1985)}]{OP85}
{Osterbrock}, D.~E. \& {Pogge}, R.~W. 1985, \apj, 297, 166

\bibitem[Osterbrock \& Shuder(1982)]{OS82} Osterbrock, 
D.~E.~\& Shuder, J.~M.\ 1982, \apjs, 49, 149 

\bibitem[Page et al.(2001)]{Pageetal01} Page, M.~J.~et al.\ 2001, 
\aap, 365, L152 


\bibitem[Petrucci et al.(2002)]{Petruccietal02} Petrucci, P.~O.~et 
al.\ 2002, \aap, 388, L5 



\bibitem[{{Pounds} {et~al.}(1995){Pounds}, {Done}, \& {Osborne}}]{PDO95}
{Pounds}, K.~A., {Done}, C., \& {Osborne}, J.~P. 1995, \mnras, 277, L5

\bibitem[Pounds et al.(2001)]{Poundsetal01} Pounds, K., Reeves, J.,
O'Brien, P., Page, K., Turner, M., \& Nayakshin, S.\ 2001, \apj, 559, 181

\bibitem[Reeves \& Turner(2000)]{RT00} Reeves, J.~N.~\& 
Turner, M.~J.~L.\ 2000, \mnras, 316, 234 

\bibitem[Reeves et al.(2001)]{Reevesetal01} Reeves, J.~N., Turner, 
M.~J.~L., Pounds, K.~A., O'Brien, P.~T., Boller, T., Ferrando, P., 
Kendziorra, E., \& Vercellone, S.\ 2001, \aap, 365, L134 


\bibitem[{{Reynolds} \& {Fabian}(1997)}]{RF97}
{Reynolds}, C.~S. \& {Fabian}, A.~C. 1997, \mnras, 290, L1

\bibitem[Reynolds(1999)]{Reynolds99} Reynolds, C.~S.\ 1999, ASP 
Conf.~Ser.~161: High Energy Processes in Accreting Black Holes, 178 


\bibitem[Reynolds(2001)]{Reynolds01} Reynolds, C.~S.\ 2001, ASP 
Conf.~Ser.~224: Probing the Physics of Active Galactic Nuclei, 105 

\bibitem[Romano et al.(2002)]{Romanoetal02} 
Romano, P., Turner, T.~J., Mathur, S., \& George, I.~M.\ 2002, \apj, 564, 
162 

\bibitem[Salvi et al.(2002)]{Salvietal02} Salvi, N.~J.~et al.\ 
2002, \mnras, 335, 177 


\bibitem[{{Ross} {et~al.}(2002){Ross}, {Fabian}, \& {Ballantyne}}]{RFB02}
{Ross}, R.~R., {Fabian}, A.~C., \& {Ballantyne}, D.~R. 2002, \mnras, in press

\bibitem[Singh, Garmire, \& Nousek(1985)]{SGN85} Singh, 
K.~P., Garmire, G.~P., \& Nousek, J.\ 1985, \apj, 297, 633 


\bibitem[Tanaka et al.(1995)]{Tanakaetal95} Tanaka, Y.~et al.\ 1995, 
\nat, 375, 659 




\bibitem[{{Turner}(1999)}]{Turneretal99}
{Turner}, et~al. 1999, in Proceedings of the 19th Texas Symposium on
  Relativistic Astrophysics and Cosmology, ed. J. Paul, T. Montmerle, \& E.
  Aubourg (Saclay: CEA), E441

\bibitem[Turner et al.(2001)]{Turneretal01} Turner, T.~J., Romano, 
P., George, I.~M., Edelson, R., Collier, S.~J., Mathur, S., \& Peterson, 
B.~M.\ 2001, \apj, 561, 131 



\bibitem[Turner et al.(2002)]{Turneretal02} Turner, T.~J.~et al.\ 
2002, \apjl, 574, L123 

\bibitem[Vaughan et al.(1999)]{Vaughanetal99} 
Vaughan, S., Reeves, J., Warwick, R., \& Edelson, R.\ 1999, \mnras, 309, 113 

\bibitem[V{\' e}ron-Cetty, V{\' e}ron, \& Gon{\c c}alves(2001)]{VVG01} V{\' e}ron-Cetty, M.-P., V{\' e}ron, P., \& Gon{\c c}alves, A.~C.\ 2001, \aap, 372, 730 



\bibitem[Wang, Zhou, \& Wang(1999)]{Wangetal99} Wang, J., Zhou, 
Y., \& Wang, T.\ 1999, \apjl, 523, L129 



\bibitem[Weaver, Gelbord, \& Yaqoob(2001)]{WGY01} Weaver, 
K.~A., Gelbord, J., \& Yaqoob, T.\ 2001, \apj, 550, 261 

\bibitem[Wilms et al.(2001)]{Wilmsetal01} Wilms, J.~;., Reynolds, 
C.~S., Begelman, M.~C., Reeves, J., Molendi, S., Staubert, R.~;., \& 
Kendziorra, E.\ 2001, \mnras, 328, L27 

\bibitem[Corbin(1991)]{Corbin91} Corbin, M.~R.\ 1991, \apjl, 
371, L51 

\bibitem[Winkler(1992)]{Winkler92} Winkler, H.\ 1992, \mnras, 
257, 677 


\bibitem[Wisotzki, Reimers, \& Wamsteker(1991)]{WWR91} 
Wisotzki, L., Reimers, D., \& Wamsteker, W.\ 1991, \aap, 247, L17 


\bibitem[Yaqoob et al.(1996)]{Yaqoobetal96} Yaqoob, T., Serlemitsos, 
P.~J., Turner, T.~J., George, I.~M., \& Nandra, K.\ 1996, \apjl, 470, L27 


\bibitem[Zheng \& O'Brien(1990)]{ZO90} Zheng, W.~\& O'Brien, 
P.~T.\ 1990, \apj, 353, 433 


\end{thebibliography}
\end{document}